\documentclass[aps,prb,preprint,superscriptaddress]{revtex4}
\usepackage{graphicx}

\begin{document}

\title{Inelastic neutron scattering due to acoustic vibrations confined
in nanoparticles: theory and experiment}

\author{Lucien Saviot}
\affiliation{Institut Carnot de Bourgogne,
UMR 5209 CNRS - Universit\'e de Bourgogne,
9 avenue A. Savary, BP 47870, F-21078 Dijon Cedex, France}
\email{lucien.saviot@u-bourgogne.fr}

\author{Caleb H. Netting}
\author{Daniel B. Murray}
\affiliation{Mathematics, Statistics and Physics Unit,
The University of British Columbia Okanagan, 3333 University Way,
Kelowna, British Columbia, Canada V1V 1V7}
\email{daniel.murray@ubc.ca}

\author{St\'ephane Rols}
\affiliation{Institut Laue Langevin,
6 rue Jules Horowitz, BP 156, 38042 Grenoble, France}

\author{Alain Mermet}
\affiliation{Universit\'e de Lyon, F-69622, Lyon, France;\\
Universit\'e Lyon 1, Villeurbanne;\\
CNRS, UMR5620, Laboratoire de Physico-Chimie des Mat\'eriaux Luminescents}

\author{Anne-Laure Papa}
\author{Catherine Pighini}
\author{Daniel Aymes}
\author{Nadine Millot}
\affiliation{Institut Carnot de Bourgogne,
UMR 5209 CNRS - Universit\'e de Bourgogne,
9 avenue A. Savary, BP 47870, F-21078 Dijon Cedex, France}

\date{\today}

\begin{abstract}

The inelastic scattering of neutrons by nanoparticles due to
acoustic vibrational modes (energy below 10 meV) confined in
nanoparticles is calculated using the Zemach-Glauber formalism.
Such vibrational modes are commonly
observed by light scattering
techniques (Brillouin or low-frequency Raman scattering).
We also report high resolution inelastic neutron scattering
measurements for anatase TiO$_2$ nanoparticles in
a loose powder.  Factors enabling the observation of such
vibrations are discussed.  These include a narrow nanoparticle
size distribution which minimizes inhomogeneous broadening of
the spectrum and the presence of hydrogen atoms oscillating with
the nanoparticle surfaces which enhances the number of scattered
neutrons.
\end{abstract}

\maketitle

\section{Introduction}

Due to the growing interest in nanoscale materials during the
last decades, the confined acoustic vibrations of nanoparticles
(NP) have been investigated both theoretically and
experimentally. Their frequency is roughly on the order of $v/d$
where $v$ is the speed of sound and $d$ is the diameter and is on
the order of 1~THz for a typical NP.

Low-frequency Raman scattering experiments can observe such
vibrations.  This can be done either for free nanoscale objects
or for NPs embedded in a matrix with large impedance
mismatch.\cite{MlayahLSS07}
Numerous Raman observations of these vibration modes are
available in the literature (see references citing
\onlinecite{Duval86}).
These same vibrations can also be observed using time resolved
pump-probe experiments\cite{BurginNL08} and far-infrared
absorption.\cite{MurrayJNO06,LiuAPL08}

In this work, we investigate the possibility of observing
confined acoustic vibrations through inelastic neutron scattering
(INS).  While neutrons interact with the nuclei of the atoms and
therefore directly probe their motion, incident photons involved
in the Raman scattering, far-infrared absorption and pump-probe
experiment processes interact with electrons.
The vibrations of
the atoms are thus probed through the electron-vibration interaction.
A distinct feature of INS is the absence of selection
rules allowing the complete vibration bands to be measured in one
experimental set. In particular, the so-called generalized density of
states (GDOS) can be measured in polycrystalline powders. In the low
frequency range, for usual bulk 3D crystals the Debye regime is observed
with a GDOS having a $\omega^2$ dependence.  There
are some past INS studies on nanoscale systems which indicated a
generally elevated background level which could be attributed to
INS from acoustic
vibrations.\cite{FultzJAP96,StuhrPRL98,BonettiJAP00,DerletPRL01,
PasquiniPRB02,YuePRL04,RolsPRL00}
However, what has been missing up until now has been the
association of a distinct spectral feature.  A key requirement
for accomplishing this is that the sample has a sufficiently
uniform collection of NPs so that the frequencies of interest have
a narrow distribution.  This enables the observation of the size
dependence of the spectral features.

\section{Experiment}

\subsection{Samples}

Unlike previous INS experiments
where observations of confined acoustic phonons were attempted,
we started from a set of samples
for which
confined acoustic modes have already been observed. TiO$_2$ nanopowders have
been prepared by continuous hydrothermal synthesis as detailed
elsewhere.\cite{PighiniJNR06} Some of us previously reported
low-frequency Raman scattering\cite{PighiniJNR06} of these samples.
Even some far infra-red features\cite{MurrayJNO06} have been tentatively
attributed to confined acoustic vibrations. The observed low-frequency
Raman peaks are narrower than the ones previously observed in the
literature\cite{IvandaJS99,MusicMSEB97,MontagnaJSGST03}
for anatase TiO$_2$ nanopowders which is in agreement with
the narrow size distribution determined by transmission electron
microscopy (TEM). The sizes determined by the broadening of the X-ray
diffraction (XRD) patterns are also in good agreement with those
determined by TEM indicating that the NPs are mostly
monocrystalline as XRD measurements are sensitive to the size of
the coherent domains.
Two different powders have been probed by INS to verify
size dependence.
The first one, referred
to as HT-5 as in Ref.~\onlinecite{PighiniJNR06}, contains NPs
with an average diameter of 3.6~nm while the second sample
HT-7 has larger NPs whose average diameter is 6.5~nm. The full width at
half maximum (FWHM) of the size distribution is close to 60\% of the
average size for both samples. These values were obtained by analyzing
the low-frequency Raman peak as in a previous work\cite{PighiniJNR06}
but using the Resonant UltraSound (RUS)
model\cite{visscher} instead of Lamb's model.\cite{lamb1882} The
agreement with the average size obtained by XRD and BET
measurements but also with the size distribution obtained from HRTEM
photos\cite{PighiniThesis} is improved.

As checked by thermogravimetry measurements, a significant amount
of species are adsorbed at the surface of the
nanopowders.\cite{LiJACS05,MuellerLangmuir03}
For our samples, typically 10\% of the total
mass could be removed from the sample by applying a low vacuum.
Further temperature annealing helps to remove more but it
also can change the NP size and their stoechiometry for
temperatures above 500K.\cite{LiJACS05}
It is well-known that anatase TiO$_2$ surfaces are stabilized by
adsorption of water molecules and formation of OH
bonds.\cite{NosakaJPCB04,BezrodnaJMS04}
Indeed, X-ray Photoemission Spectroscopy (XPS) measurements
(which require ultra-high vacuum) show that for sample HT-7, more
than $10\%$ of the total number of atoms are hydrogen
atoms.\cite{PighiniThesis}
The neutron scattering cross
section of a single hydrogen atom is roughly 20 times more
than that of a single oxygen or titanium atom.  Therefore, in
this ultra-high vacuum condition
the contribution of hydrogen atoms to the total
scattering would be more than $\frac23$. This value is the worst
case estimate of the minimum contribution expected in this work
as the proportion of hydrogen atoms increases when decreasing the
size but also when more water molecules are adsorbed
as is the case during INS measurements because
the vacuum is not as good.

Due to the high scattering cross-section of hydrogen compared to
oxygen and titanium, it is not clear if neutron scattering from
hydrogen could hinder the observation of the NPs
vibrational density of states as in a recent work\cite{LevchenkoJPCA07}
or instead enhance it if hydrogen atoms move with the
NPs' surface.  Therefore some heat treatments were
performed to check the influence of adsorbed water.

\subsection{Measurements}

INS experiments have been performed with the IN6 time-focusing time of
flight spectrometer at the Institut Laue-Langevin (ILL). For the first
experiments, the neutron wavelength was 5.1 \AA. For each sample, 500~mg
of powder were wrapped inside a non-hermetic aluminum foil and put inside
a cryofurnace. The empty foil signal was removed and the efficiencies of
the detector were corrected using a vanadium sample run. Measurements
were carried out first at 300K, then the samples were heated at 500K and
annealed for a few hours. During annealing, the helium atmosphere
surrounding the sample was purged a few times to removed desorbed water.
Then further measurements were performed at 300, 200, 100 and 10K to
check for a quasielastic signal.

Another round of experiments was performed later in order to have
better statistics. The neutron wavelength was 4.1 \AA, the
temperature was 500K and the accumulation time much longer. The
aluminium foil was inside a cryoloop. In this configuration, the foil
was more hermetic allowing to have more water molecules adsorbed at the
surface of the NPs.

\subsection{Results}

\begin{figure}
 \includegraphics[width=\columnwidth]{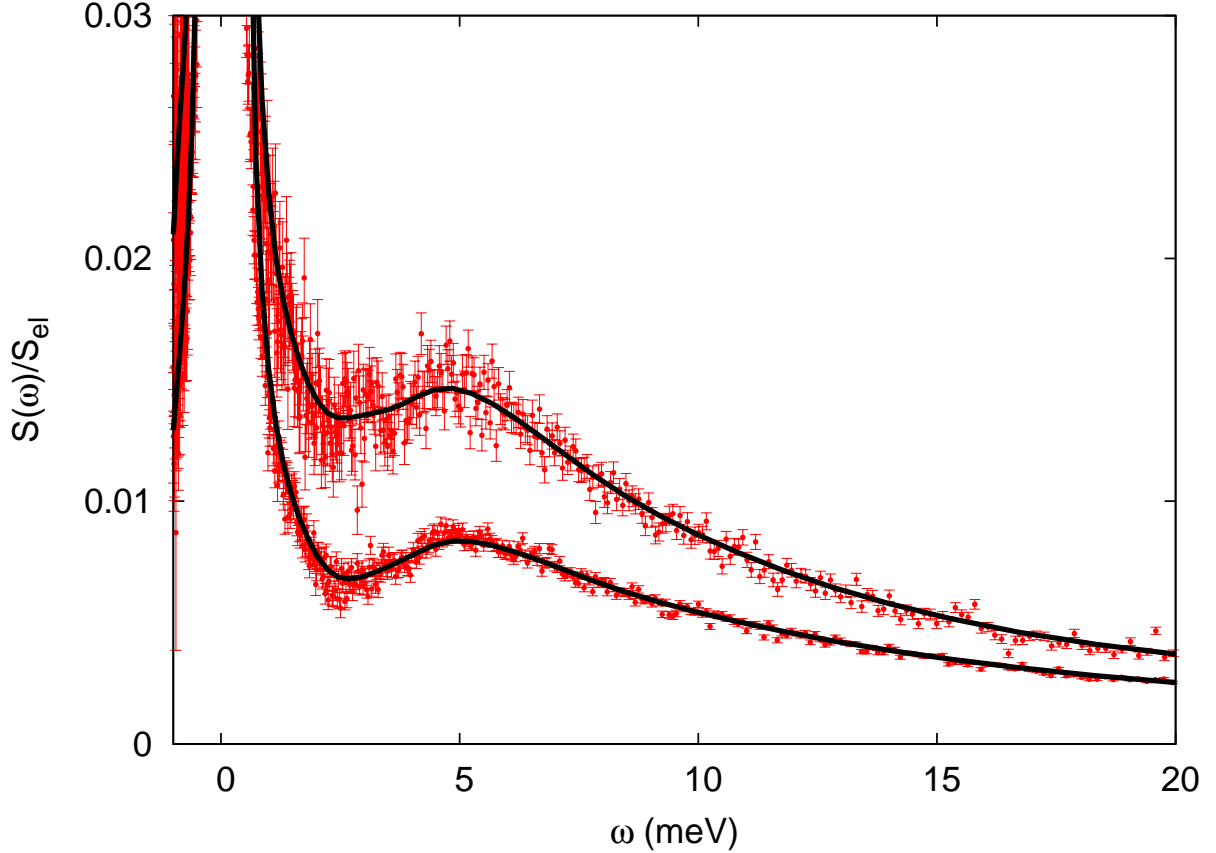}
 \caption{\label{SwHT5R1} (Color online) $S(\omega)/S_{el}$ with experimental
 error bars for an anatase TiO$_2$ nanopowder with nanoparticles having
 an average diameter $d$=3.6~nm (sample HT5).
 The neutron wavelength is 5.1~\AA{} and the temperature is
 T=300K before (top) and after (bottom) the heat treatment and
 atmosphere replacement at 500K.
 The black lines obtained by smoothing the data are only a guide for
 the eye.}
\end{figure}

The incoherent neutron scattering functions $S(\omega)$
measured on
sample HT5 at 300K before and after heat treatment at 500K are presented
in Fig.~\ref{SwHT5R1}.
It was obtained by summing $S(\theta,\omega)$ over all available
detector angles after checking that it doesn't depend on the scattering
wavevector.
These spectra have been normalized by the area under the elastic
peak ($S_{el}$) obtained by fitting by a Gaussian.
Error bars are also drawn representing the experimental error. This
error is obtained by propagating through all the calculations the error
(square root of the number of neutrons) for each detector and time channel.
The significant decrease of the inelastic intensity is
clearly related to the desorption of water which occurred during the heat
treatment and also because the atmosphere was purged during this
process. As a result, there are fewer hydrogen atoms at the surface
of the TiO$_2$ NPs after the heat treatment. A peak is clearly
identified close to 5~meV and its position is the same
before and after the heat treatment.

\begin{figure}
 \includegraphics[width=\columnwidth]{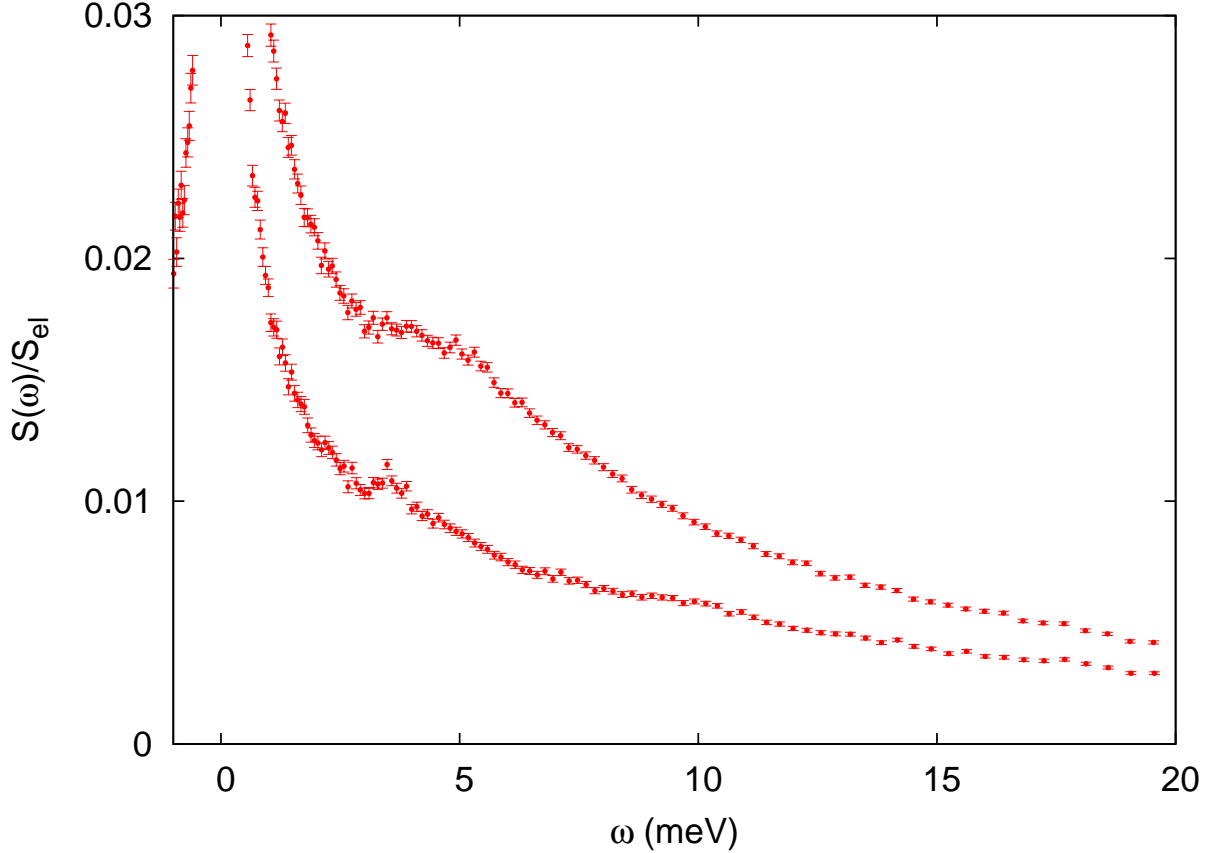}
 \caption{\label{Sw500} (Color online) $S(\omega)/S_{el}$
 with experimental error bars
 for two anatase TiO$_2$ nanopowders with nanoparticles having
 an average diameter $d$=3.6~nm (sample HT5, top)
 and $d$=6.5~nm (sample HT7, bottom).
 The neutron wavelength is 4.1~\AA{} and the temperature T=500K.}
\end{figure}

Similar experiments were performed for sample HT7 but the signal
measured was smaller preventing the observation of a similar
low-frequency peak. This signal could be improved by increasing the
temperature and keeping more hydrogen at the surface in the second
round of experiments at the expense of an increased quasielastic
signal. Figure \ref{Sw500} presents the measured $S(\omega)$
for samples HT5 and HT7 at 500K.
The shape of the spectrum clearly depends on the size of the
nanoparticles as the peak observed for sample HT5 is not present
for sample HT7. This change is compatible with a shift of the peak
towards lower energies as the radius of the nanoparticle increases.

\section{Theory}

\subsection{Inelastic neutron scattering}

There is no theoretical formalism presently available which
is developed specifically for INS
from a NP powder sample.  Since the nanopowder
used in this work is of low density, it is a reasonable
approximation to treat the constituent NPs as
mechanically independent,
each NP vibrating as if free.
Thus, the INS due to a single free NP can be
theoretically predicted
and the result applied to the entire sample, taking into
consideration the size distribution of the nanopowder.
The total neutron scattering cross section is simply
additive.
Under our free NP approximation,
there is no coherence between the mechanical
vibrations of different NPs.

Notations are the same as in Ref.~\onlinecite{HudsonJCP74}.
Consider a single NP consisting of atoms which
are labeled by integer index $\nu$.  Atom $\nu$ has mass
$m_{\nu}$ and scattering length $a_{\nu}$.
As explained in Ref.~\onlinecite{HudsonJCP74},
$a^2 = 1/(4\pi) (\sigma_{coh} + \sigma_{inc})$
where $\sigma_{coh}$ and $\sigma_{inc}$ are, respectively,
the coherent and incoherent bound cross sections for the
given type of atom.
The NP
has vibrational modes indexed by $\lambda$, with frequencies
$\omega_{\lambda}$ and mass-weighted
mode displacements $\vec{e}_\nu^{\,(\lambda)}$.
A general vibrational state transition is indicated by a set of
integers $[n_{\lambda}]$ so that the energy gain of the neutron
after the scattering is $\hbar \omega$ where

\begin{equation}
\omega = -\sum_{\lambda} n_{\lambda} \omega_{\lambda}
\end{equation}

The initial wavevector of the neutron is $\vec{k}_o$ with
magnitude $k_o$.  The final wavevector is $\vec{k}$ with
magnitude $k$.  The change of wavevector is
$\vec\kappa = \vec k - \vec {k_o}$.

Using previous theories on the differential cross section of
INS due to vibrational modes
of a molecule\cite{ZemachPR56,HudsonJCP74},
a formula for the neutron scattering cross section due to a
given general vibrational state transition of a molecule can be
written as:
\begin{equation}
<\sigma([n_\lambda],\theta)>_T = \frac{k}{k_o} \sum_\nu a_\nu^2 W_\nu
\end{equation}
where
\begin{eqnarray}
\label{eqW}
\nonumber
W_\nu & = & \prod_\lambda
\exp \left[-
               \left( \frac{\hbar}{2 m_\nu \omega_\lambda} \right)
               \left( \vec{\kappa} \cdot \vec{e}_\nu^{\,(\lambda)}   \right)^2
               \coth\left( \frac{\hbar \omega_\lambda}{2 k_B T}  \right)
            \right]
\\
\nonumber
& & \times \exp \left( \frac{-n_\lambda \hbar \omega_\lambda }{2 k_B T} \right)
\\
& & \times I_{n_\lambda} \left(
    \frac{\left(\hbar/2 m_\nu \omega_\lambda\right)
          \left( \vec{\kappa} \cdot \vec{e}_\nu^{\,(\lambda)}   \right)^2}
         {\sinh( \hbar \omega_\lambda / 2 k_B T ) }
                     \right)
\end{eqnarray}
and $I_{n_\lambda}$ is a modified Bessel function of order
$n_\lambda$.

This expression ignores off-diagonal contributions to $\sigma$
which come from coherent interference between different atoms.
The formula is valid whether the neutron scattering is coherent
or incoherent.  As in Hudson et al.\cite{HudsonJCP74}, the
contribution to the total cross section from the off-diagonal
terms is expected to be negligible.  In the following, the
coherent scattering terms are omitted because as discussed before
most of the experimental scattering events are due to the
incoherent scattering by hydrogen atoms.  However, we expect the
motion of these atoms to be representative of the motion of the
surface of the NP.  This is true for the OH bonds at the
surface because the acoustic phonon frequencies are very small
compared to the frequency of the OH bond.  The situation for water
molecules physisorbed on the surface is of course more complex.
However the atmosphere was purged, the helium atmosphere pressure
was around 10~mbar and the temperature was increased in order
to reduce the amount of such molecules.  As a result, it is safe
to neglect the contribution of these molecules.

In contrast to many INS observations of vibrational modes of
molecules, the above expressions can be substantially simplified
for the experimental situation we are considering because of the
large mass of the NP and also the long wavelength of the incident
neutrons (which makes $k_o$ small).  For both the first and third
factors in Eq.~(\ref{eqW}), the function argument is on the order of
$k^2 k_B T / \omega^2 M$, where $M$ is the mass of a NP.  At room
temperature, this is approximately 0.00002,
so linear approximations of the functions can thus be used
and higher
corrections are thus negligible.  The first factor is the
Debye-Waller factor. It is well-approximated as 1 in every
situation relevant to our experiment.
This is partly a result of the smallness of $\kappa$
for this experimental setup.  It is also related to the low
frequency involved.
For the third factor, we
make use of the limiting form of the modified Bessel function
$I_n(x) \simeq  (x/2)^{|n|} / ( |n|! )$.  Since $x$ is so small,
only terms with $n$ = -1 or +1 need to be included.  In our
experiment, we only collect data for increasing neutron energy,
which corresponds to $n_\lambda$=1.  We are now only talking
about a single mode with frequency $\omega$, so we no longer
mention $\lambda$.  Then,
\begin{equation}
W_\nu =
\frac{\hbar}{4 m_\nu \omega}
\exp \left( \frac{- \hbar \omega }{2 k_B T} \right)
    \frac{( \vec{\kappa} \cdot \vec{e}_\nu)^2}
         {\sinh{\frac{\hbar \omega}{2 k_B T}} }
\end{equation}

The next step is to consider the random orientation of the
NPs.  As a result, the orientation of the motion of a
given atom is random with respect to the direction of
$\vec{\kappa}$.  Thus, $( \vec{\kappa} \cdot \vec{e}_\nu)^2$
can be replaced by $\frac{1}{3} \kappa^2 e_\nu^2$.

To evaluate the total inelastic cross section for this
mode (to be called $\sigma_{inel}$) we have to integrate the
differential cross section $\sigma(\theta)$ over all possible
final directions of the final neutron wavevector $\vec{k}$.
$\theta$ being the angle between $\vec k$ and $\vec {k_o}$,
\begin{eqnarray}
 \kappa^2   & = & k^2 +k_o^2 -2 k k_o \cos{\theta}
\end{eqnarray}

Therefore
\begin{equation}
 <\kappa^2> = k^2 +k_o^2 -2 k k_o <\cos{\theta}>
\end{equation}

This expression depends on $\omega$ as $k_o$ is fixed by the
experimental conditions,
$k=\sqrt{k_o^2+\frac{2 m_n \omega}{\hbar}}$
and $<\cos{\theta}>$ can be calculated from the geometry of the
experimental setup.
Here, we took $<\cos{\theta}>$ = 0.2804.

The final result is a nearly exact expression for the inelastic cross
section:
\begin{eqnarray}
\nonumber
\sigma_{inel} = k^2  \left(\frac{k}{k_o} + \frac{k_o}{k}
-2<\cos{\theta}>\right)
\\
\nonumber
\times
\exp \left( \frac{- \hbar \omega }{2 k_B T} \right)
\frac{1}{12}
\frac{\hbar}{\omega}
\frac{1}{\sinh( \hbar \omega / 2 k_B T)}
\\
\times
\sum_\nu
\left( \frac{\sigma_\nu e_\nu^2}{m_\nu} \right)
\end{eqnarray}

\subsection{Acoustic modes and size distribution}

Let us first figure out the summation over the different atoms
($\nu$) of $\frac{\sigma_\nu e_\nu^2}{m_\nu}$. Summations over
all modes as well as the size distribution of NPs in
order to get a quantity closer to the experimentally INS
spectrum will be considered later.
Let $\sigma_H$ and $m_H$ respectively denote the neutron
scattering cross section and mass of a hydrogen atom.
Assuming that only the scattering by H surface
atoms matters and neglecting the scattering by other atoms (Ti
and O), we can rewrite this summation as being over all the
H atoms $\frac{\sigma_H}{m_H} \sum_{\nu_H} e_{\nu_H}^2$.
$\rho$ is the density of the NP and $V$ is the volume.
$N_H$ is the number of hydrogen atoms fixed to the surface
of the NP.
Introducing the mean square displacement at the surface U2S defined as
the ratio of the surface and volume average of the square displacement
as in Ref.~\onlinecite{SaviotPRB05}, it follows that:
\begin{eqnarray}
\nonumber
  \sum_\nu \left( \frac{\sigma_\nu e_\nu^2}{m_\nu} \right)
  & = &
  \frac{\sigma_H}{\rho} \frac{N_H}{V} \mathrm{U2S}\\
  & = &
  \frac{\sigma_H d_H}{\rho} \frac3R \mathrm{U2S}
\end{eqnarray}
where $d_H$ is the surface density of hydrogen atoms which we
will suppose to be independent of $R$ in the following.  The
summation above is over the hydrogen atoms only.  This expression
already shows that vibrational modes having a large amplitude at
the surface and larger NPs will scatter neutrons more.

As with most low-frequency Raman interpretations, we model the
nanopowder as an ensemble of spherical elastic NPs.
The spherical shape is consistent with the average shape observed
by TEM.  All the frequencies of homogeneous elastically
anisotropic spheres
$\omega_\lambda(R)=\frac{\alpha_\lambda}{R}$ can be calculated
using the method introduced by Visscher et al.\cite{visscher}
Taking into account the degeneracy of the different modes and
the radius distribution $P(R)$, the incoherent neutron scattering
function can be calculated. For an ideal system consisting of an
ensemble of identical NPs, it would be non-zero only
for the discrete set of frequencies of the NPs.  For a
real system, these discrete frequencies are broadened and the
maxima are shifted by the distribution of NPs.

For a single vibration mode and a single particle, we obtain the
following expression for incoherent
INS
due to a uniform distribution of H atoms on the surface of the
nanosphere:
\begin{equation}
\sigma_{inel}(\omega_\lambda) \sim
\frac1{\alpha_\lambda}
\frac{\kappa^2}{k_o k}
\frac{\mathrm{U2S}_\lambda}
{\exp\left( \frac{\hbar \omega_\lambda}{k_B T}\right)-1}
\end{equation}
where only the terms depending on $R$ or $\lambda$ have been retained.
It should be noted that $\mathrm{U2S}_\lambda$ does not depend on $R$.

Now summing over the different modes $\lambda$ and radii $R$ we
obtain:
\begin{equation}
 S(\omega) \sim
 \sum_\lambda  \mathrm{U2S}_\lambda
\int_R \frac{\kappa^2}{k_o k}
\frac{P(R) \delta\left(\omega-\omega_\lambda(R)\right) dR}
{\alpha_\lambda (\exp\left( \frac{\hbar \omega}{k_B T}\right)-1)}
\end{equation}

Using the properties of the $\delta$ function, this expression
translates into:
\begin{equation}
 S(\omega) \sim
\frac{\kappa^2}{k_o k}
\frac
{\sum_\lambda \mathrm{U2S}_\lambda P(\alpha_\lambda/\omega)}
{\omega^3 \left(\exp\left( \frac{\hbar \omega}{k_B T}\right)-1\right)}
\label{final}
\end{equation}

\begin{figure}
 \includegraphics[width=\columnwidth]{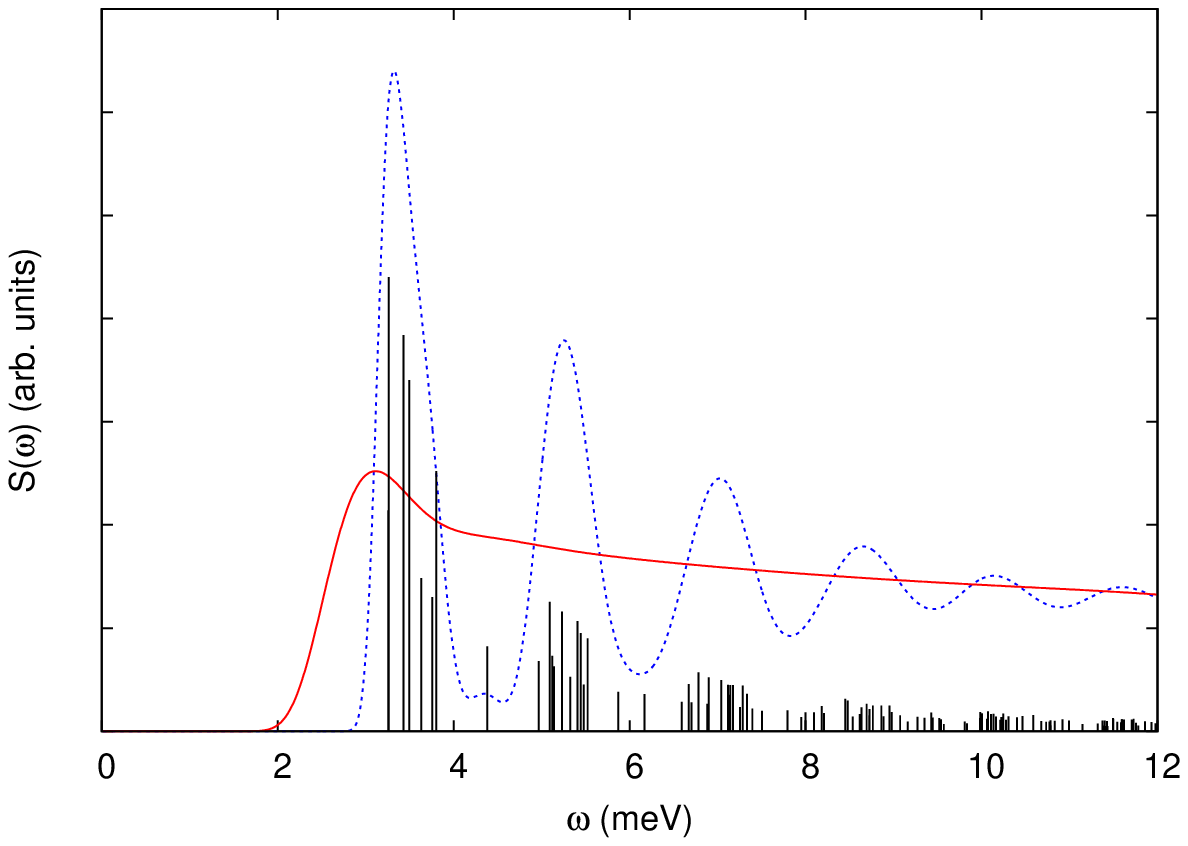}
 \caption{\label{calc} (Color online) $S(\omega)$ calculated for sample HT5 with
 varying Gaussian size distribution.
 The average diameter is $d$=3.6~nm in all cases.
 The FWHM of the size
 distribution is 10\% and 50\% of the average size for the dotted and
 continuous curves respectively.
 The neutron wavelength is 5.1~\AA. The
 vertical lines at the bottom show the theoretical frequencies of the
 eigenvibrations for a spherical particle having a 3.6~nm diameter.
 The height of these lines is proportional to the degeneracy of the
 modes (either 1 or 2) multiplied by URS$/\omega^3$.}
\end{figure}

By using the RUS algorithm\cite{visscher} and the anisotropic
elastic parameters for anatase TiO$_2$ from
Ref.~\onlinecite{IugaEPJ07} we could calculate the
$\omega_\lambda$ for nanospheres, the associated displacements
and their classification according to their irreducible
representation.
This approach relies on continuum elasticity.
The shape and elastic tensor for a given object are defined.
Then the displacements of the eigenmodes are expanded on a $x^i y^j
z^k$ basis. For the free boundary condition, the eigensolution problem
is turned into a real generalized symmetric-definite eigenproblem
through the use of the Hamilton's principle. Such a problem is
efficiently
solved on any modern computer.
Using this method, we could
calculate the $\mathrm{U2S}_\lambda$ by numerical integration.
Some values for the first few modes are presented in Tab.~\ref{RUS}.
Calculations for sample HT5 could then be performed and are
presented in Fig.~\ref{calc} using the average size for sample HT-5
and different Gaussian size distributions.
For comparison with the frequently used isotropic model, the anisotropic
eigenmodes obtained with the RUS algorithm have been projected onto the
Lamb's modes and the main contributions are shown in the last column
of Tab.~\ref{RUS}. This demonstrates the splitting and mixing of the
isotropic modes due to the anisotropy of the tetragonal elasticity
similarly to what has been recently reported for the cubic
symmetry.\cite{PortalesPNAS08}

\begin{table}[ht!]
 \begin{tabular}{|c|c|c|c|}
 \hline
$\omega$ (meV) & U2S & IR & projection onto Lamb's modes\\
\hline
3.26 & 0.9 & A$_{1g}$ & (SPH,$\ell=2$)\\
3.26 & 1.9 & E$_{u}$  & (TOR,$\ell=2$)\\
3.43 & 1.9 & A$_{1u}$ & (TOR,$\ell=2$)\\
3.49 & 1.8 & B$_{1u}$ & (TOR,$\ell=2$)\\
3.63 & 0.9 & E$_{g}$  & (SPH,$\ell=2$)\\
3.76 & 0.9 & B$_{2g}$ & (SPH,$\ell=2$)\\
3.80 & 1.7 & B$_{2u}$ & (TOR,$\ell=2$)\\
4.38 & 0.9 & B$_{1g}$ & (SPH,$\ell=2$)\\
4.97 & 1.0 & A$_{2u}$ & (SPH,$\ell=1$)+(SPH,$\ell=3$)\\
5.09 & 2.0 & B$_{2g}$ & (TOR,$\ell=3$)\\
5.12 & 1.2 & E$_{u}$  & (SPH,$\ell=1$)+(SPH,$\ell=3$)\\
5.14 & 1.1 & A$_{2u}$ & (SPH,$\ell=3$)+(SPH,$\ell=1$)\\
5.23 & 2.0 & E$_{g}$  & (TOR,$\ell=3$)\\
5.32 & 1.0 & E$_{u}$  & (SPH,$\ell=3$)\\
5.41 & 2.1 & A$_{2g}$ & (TOR,$\ell=3$)\\
\vdots & \vdots & \vdots & \vdots\\
\hline
 \end{tabular}
\caption{\label{RUS}Frequencies, U2S and irreducible
representations (IR) for the vibration modes of a $d$=3.6~nm
anatase TiO$_2$ NP. Degeneracy is 1 except for E modes for
which it is 2. The last column show the correspondence between
the spheroidal (SPH) and torsional (TOR) Lamb's modes having an
angular momentum $\ell$ of an isotropic sphere.}
\end{table}

\section{Discussion}
\subsection{General agreement and considerations regarding the vibrations}
As shown by Fig.~\ref{calc}, the discreteness of the acoustic
vibrations in NPs is expected to be observed in the INS
spectrum in a similar way that such vibrations can be observed
by inelastic light scattering experiments.  Of course, the size
distribution plays a significant role and may hide the narrow features
expected for a single NP. This effect is more dramatic for
INS because of the absence of selection rules which makes all vibration
modes observable.

Apart from the elastic and quasielastic contributions which are
not taken into account in our calculations, the calculated
$S(\omega)$ in Fig.~\ref{calc} is qualitatively in agreement with
the experimental spectra for sample HT-5 in Fig.~\ref{SwHT5R1}
and \ref{Sw500}.  The main feature is a peak around 3~meV
resulting mainly from the lowest frequency vibrations and the
frequency gap in the long wavelength regime,
extending from 0 up to about 2 meV.
The position of the experimental peak for this sample is close to 5~meV
to be compared with the calculated position at 3~meV. This discrepancy
will be discussed in the following section.
The other features at higher
frequencies are progressively smoothed when increasing the width
of the size distribution because the frequency gaps are too
narrow.

Instead of using the numerical RUS approach to calculate the NP
vibration modes and frequencies, it would have been simpler to
use the isotropic approximation (Lamb's model\cite{lamb1882})
for which an exact solution exists. However, using such a model
would have resulted in artificially narrow peaks in the INS
spectrum because the vibration mode degeneracy is higher in that
case. A single domain spherical anatase TiO$_2$ nanocrystal has
tetragonal symmetry and the isotropic vibrations are split into
modes having degeneracy at most two for this symmetry.  Therefore
not using the isotropic approximation enables us to have more
accurate calculations of the frequencies and U2S but also
prevents the appearance of some artifacts due to unrealistic
degeneracies of the vibrations.  For example, the spheroidal and
torsional modes with $\ell = 2$ for an isotropic system split
into the first (A$_{1g}$, B$_{1g}$, B$_{2g}$, E$_{g}$)
and (A$_{1u}$, B$_{1u}$, B$_{2u}$, E$_{u}$) modes respectively.
The frequency splitting due to the tetragonal symmetry is quite
significant.  Tab.~\ref{RUS} shows that due to the U2S factor,
some modes will scatter neutrons significantly more efficiently
than some others.  The U2S values for the torsional modes are
roughly two times larger than that for the spheroidal modes
making INS more sensitive to torsional modes. Because of the
$1/\omega^3$ factor in Eq.~\ref{final}, the lowest frequency
modes contribute more to $S(\omega)$.
This is clearly evidenced by the variation of the heights of the
vertical lines in Fig.~\ref{calc}.
Therefore, the main
feature in the incoherent neutron scattering function comes
mainly from the scattering by the lowest (A$_{1u}$, B$_{1u}$,
B$_{2u}$, E$_{u}$) modes (resulting from the splitting of the
isotropic torsional mode with $\ell=2$).

Being based on a continuum description, the RUS model is
applicable only in a limited frequency range.  Optical phonons of
anatase TiO$_2$ exist down to approximately 17.8~meV so this
model cannot be trusted above this value.  More importantly, the
summation in Eq.~\ref{final} was performed for a limited range of
$\lambda$ (the 500 lowest frequencies) which results in an even
smaller frequency range where the summation converged.  For
sample HT5 and the size distributions used in Fig.~\ref{calc},
the convergence was reached for $\omega < 12$~meV.

\subsection{Explanations for the discrepancy between experiment and
theory}

In the following, we will discuss several limitations in the
model proposed in this work which could explain the discrepancy
between the experimental results and the calculations. Regarding the
vibrations, the applicability of continuum models is questionable
especially for the modes at high frequencies or for very small
sizes. However, based on comparison with existing atomistic
calculations\cite{ChengPRB05,ErratumChengPRB05,CombePRB07,RamirezJASA08},
continuum approaches are valid for the sizes considered in this
work at least up to the first breathing mode. In the present case,
the frequency for this mode is more than 2.5 times the lowest
eigenfrequency. Therefore it is reasonable to expect the present
description of the vibrations to be sufficient to reproduce the main
peak of the INS spectrum.

Another source of discrepancy could be the shape and crystalline
structure of the NPs. In the proposed model, the NPs
are perfectly spherical single domain nanocrystals. HRTEM
photos\cite{PighiniThesis} confirm that this is
at least a reasonable approximation.
However the variations of shape or crystallinity required to shift position
of the first calculated peak to the one of the experimental spectra is
too large for this to be the most important reason. The same HRTEM
photos also rule out the presence of a significant amount of amorphous
TiO$_2$ which could have dominated the spectrum.

Another reason for the discrepancy could be
the assumption that the surface
density of hydrogen atoms does not depend on the NP size. In order to
better fit the calculated $S(\omega)$ to the experimental spectra, the
surface density would have to increase with decreasing size to shift the
calculated first peak towards higher energies. Such a variation would be
in agreement with a higher density of OH groups close to surface
discontinuities. The surface of larger NPs should contain less
discontinuities provided they can form larger more stable flat faces. In
this case however, the variation of $d_H$ with $R$ would have to be
very large to obtain a better agreement.

\subsection{Common features of INS from nanopowders and glasses}

The presently reported data for TiO$_2$ NP powders share
obvious similarities with the Boson Peak (BP) modes observed from
glasses: in the acoustic regime, the INS curves evidence low frequency
excitations that significantly contrast with the monotonic behaviour
of the Debye density of states from extended crystalline solids.
Furthermore, as in glasses, these harmonic excitations
coexist with
a quasielastic signal that strongly affects the visibility of the
inelastic one at high temperature (Fig.~\ref{Sw500}). As in many porous
systems like carbon nanotubes, zeolites or silica gels, the quasielastic
signal observed in the INS spectra of TiO$_2$ NP
powders most
likely arises from relaxational motions of the hydrogen atoms associated
with residual adsorbed water or TiOH bonds.

As shown above, the low frequency excitations observed in the case
of a disordered assembly of NPs
essentially identify with
their lowest frequency fundamental modes. Qualitatively, the inelastic
bump in Fig.~\ref{SwHT5R1} reflects the low frequency cutoff in
the vibrational density of states, below which no vibration mode
can be accommodated inside the NP (see frequency gap in
Fig.~\ref{calc}). Interestingly, such a finite size
cutoff has been
proposed to explain the BP origin, regardless of its spectral details
that vary according the nature of the glass.\cite{Duval1990} The basic
idea behind this interpretation is that the BP excitations are similar
to NP-like modes of nanometric inhomogeneities (nanodomains)
that are natively formed within the glass. The main difference between a
NP assembly and a glassy nanostructure is that in the latter
case the low frequency excitations are expected to be less well defined
due to poor elastic contrast with the surrounding disorder (leading to
hybridization with delocalized acoustic modes\cite{Duval2007}) and also
expected poor nanodomain shape definition. Nevertheless, as shown from
numerical simulations,\cite{Leonforte2005} the elastic disruption of the
glass network at the nanometer scale suffices to generate low frequency
excitations that the longstanding ``continuous random network'' glass
model structure is unable to account for. The present INS study provides
further indication that the glass BP is intrinsically related to a
specific nano-texture, in line with previous INS studies of zeolite
amorphization.\cite{Greaves2005}

\section{Conclusion}

The experimental and theoretical results presented in this paper
demonstrate the possibility of using INS to observe acoustic modes
confined in NPs.
The experimental peaks we observed for one sample are in the right
frequency range. The absence of a clear peak for the other sample
with larger nanoparticles is compatible with a peak shifted towards
lower frequencies as expected from our model and mostly masked by the
quasielastic signal.
These measurements were facilitated
by the presence of hydrogen atoms due to OH groups at the surface.
Because of the large incoherent scattering cross-section of these
atoms and their increasing number with respect to the total number of
atoms for decreasing NP sizes, their contribution to the INS spectra
dominates for very small sizes. For larger sizes, a correct treatment
would require taking into account the scattering by volume atoms
which is beyond the scope of this paper. Unlike light scattering
measurements, the lack of selection rules for INS enables the
observation of all the vibration modes. Therefore INS measurements
on samples with narrower size distributions would provide valuable
experimental data to compare to current models of the vibrations of
NPs.

\bibliography{instio2}
\end{document}